\documentclass[sigconf]{acmart}
\usepackage{enumitem}
\usepackage{bm}
\usepackage{cuted}
\usepackage{float}
\usepackage[ruled,linesnumbered]{algorithm2e}

\usepackage{enumitem}
\usepackage[linesnumbered,ruled]{algorithm2e}
\usepackage{colortbl}
\usepackage{multirow}
\usepackage{graphicx}
\usepackage[normalem]{ulem}
\useunder{\uline}{\ul}{}

\definecolor{gblue}{RGB}{66,133,244}

\newcommand{\emphcolorbluee}[1]{\colorbox{gblue!10}{#1}}

\newcommand{\emphcolorbrownn}[1]{\colorbox{brown!15}{#1}}

\newcommand{\emphcolorgreenn}[1]{\colorbox{green!8}{#1}}

\newcommand{\emphcoloryelloww}[1]{\colorbox{yellow!8}{#1}}
\usepackage{arydshln}
\makeatletter
\def\adl@drawiv#1#2#3{%
        \hskip.5\tabcolsep
        \xleaders#3{#2.5\@tempdimb #1{1}#2.5\@tempdimb}%
                #2\z@ plus1fil minus1fil\relax
        \hskip.5\tabcolsep}
\newcommand{\cdashlinelr}[1]{%
  \noalign{\vskip\aboverulesep
           \global\let\@dashdrawstore\adl@draw
           \global\let\adl@draw\adl@drawiv}
  \cdashline{#1}
  \noalign{\global\let\adl@draw\@dashdrawstore
           \vskip\belowrulesep}}
\makeatother

\usepackage[normalem]{ulem} % gcm: 下划线:\sout{xxxx}

\usepackage{colortbl} % 引入colortbl包
\definecolor{mygray}{rgb}{0.9, 0.9, 0.9}
\definecolor{myred}{rgb}{0.68627451, 0.14117647, 0.09803922}
\AtBeginDocument{%
  }

\setcopyright{acmlicensed}
\copyrightyear{2018}
\acmYear{2018}
\acmDOI{XXXXXXX.XXXXXXX}
\acmConference[Conference acronym 'XX]{Make sure to enter the correct
  conference title from your rights confirmation email}{June 03--05,
  2018}{Woodstock, NY}
\acmISBN{978-1-4503-XXXX-X/2018/06}

\begin{document}

%%
%% The "title" command has an optional parameter,
%% allowing the author to define a "short title" to be used in page headers.
% \title{COMPASS: Leveraging Collaboration Preference Alignment for Generative Recommender}
% \title{Practice on Unifying Learning to Rank with Generative Recommender at Kuaishou}

% \title{You Only Recommend Once: Unifying Retrieve and Rank with Generative Recommender and Preference Alignment}
\title{OneRec: Unifying Retrieve and Rank with Generative Recommender and Preference Alignment}
%%
%% The "author" command and its associated commands are used to define
%% the authors and their affiliations.
%% Of note is the shared affiliation of the first two authors, and the
%% "authornote" and "authornotemark" commands
%% used to denote shared contribution to the research.
\author{Jiaxin Deng}
\authornote{Equal contribution.}
\affiliation{%
  % \institution{State Key Laboratory of Multimodal Artifcial Intelligence Systems, Institute of Automation, Chinese Academy of Sciences}
  \institution{KuaiShou Inc.}
  \country{Beijing, China}
}
 \email{dengjiaxin03@kuaishou.com}
\author{Shiyao Wang}
\authornotemark[1]
\affiliation{%
  \institution{KuaiShou Inc.}
   \country{Beijing, China}
   }
   \email{wangshiyao08@kuaishou.com}
\author{Kuo Cai}
\authornotemark[1]
\affiliation{%
  \institution{KuaiShou Inc.}
   \country{Beijing, China}
   }
   \email{caikuo@kuaishou.com}
\author{Lejian Ren}
\authornotemark[1]
\affiliation{%
  \institution{KuaiShou Inc.}
   \country{Beijing, China}
   }
   \email{renlejian@kuaishou.com}
\author{Qigen Hu}
\authornotemark[1]
% \authornotemark[1]
\affiliation{%
  \institution{KuaiShou Inc.}
   \country{Beijing, China}
   }
   \email{huqigen03@kuaishou.com}
\author{Weifeng Ding}
\authornotemark[1]
% \authornotemark[1]
\affiliation{%
  \institution{KuaiShou Inc.}
   \country{Beijing, China}
   }
   \email{dingweifeng@kuaishou.com}
\author{Qiang Luo}
\authornotemark[1]
\affiliation{%
  \institution{KuaiShou Inc.}
   \country{Beijing, China}
   }
   \email{luoqiang@kuaishou.com}
\author{Guorui Zhou}
\authornotemark[1]
\authornote{Corresponding author.}
\affiliation{%
  \institution{KuaiShou Inc.}
   \country{Beijing, China}
   }
   \email{zhouguorui@kuaishou.com}

%%
%% By default, the full list of authors will be used in the page
%% headers. Often, this list is too long, and will overlap
%% other information printed in the page headers. This command allows
%% the author to define a more concise list
%% of authors' names for this purpose.
\renewcommand{\shortauthors}{Deng et al.}

%%
%% The abstract is a short summary of the work to be presented in the
%% article.
\begin{abstract}
  Recently, generative retrieval-based recommendation systems (GRs) have emerged as a promising paradigm by directly generating candidate videos in an autoregressive manner. However, most modern recommender systems adopt a retrieve-and-rank strategy, where the generative model functions only as a selector during the retrieval stage. In this paper, we propose \textbf{OneRec}, which replaces the cascaded learning framework with a unified generative model. \textit{To the best of our knowledge, this is the first end-to-end generative model that significantly surpasses current complex and well-designed recommender systems in real-world scenarios}.
Specifically, OneRec includes: 1) \textbf{an encoder-decoder structure}, which encodes the user's historical behavior sequences and gradually decodes the videos that the user may be interested in. We adopt sparse Mixture-of-Experts (MoE) to scale model capacity without proportionally increasing computational FLOPs.
2) \textbf{a session-wise generation approach}. In contrast to traditional next-item prediction, we propose a session-wise generation, which is more elegant and contextually coherent than point-by-point generation that relies on hand-crafted rules to properly combine the generated results. 3) \textbf{an Iterative Preference Alignment module} combined with Direct Preference Optimization (DPO) to enhance the quality of the generated results. Unlike DPO in NLP, a recommendation system typically has only one opportunity to display results for each user's browsing request, making it impossible to obtain positive and negative samples simultaneously. To address this limitation, We design a reward model to simulate user generation and customize the sampling strategy according to the attributes of the recommendation system's online learning.
Extensive experiments have demonstrated that a limited number of DPO samples can align user interest preferences and significantly improve the quality of generated results. We deployed OneRec in the main scene of Kuaishou, a short video recommendation platform with hundreds of millions of daily active users, achieving a 1.6\% increase in watch-time, which is a substantial improvement.

\end{abstract}

%%
%% The code below is generated by the tool at http://dl.acm.org/ccs.cfm.
%% Please copy and paste the code instead of the example below.
%%


%%
%% Keywords. The author(s) should pick words that accurately describe
%% the work being presented. Separate the keywords with commas.
%% A "teaser" image appears between the author and affiliation
%% information and the body of the document, and typically spans the
%% page.
% \begin{teaserfigure}
%   \includegraphics[width=\textwidth]{sampleteaser}
%   \caption{Seattle Mariners at Spring Training, 2010.}
%   \Description{Enjoying the baseball game from the third-base
%   seats. Ichiro Suzuki preparing to bat.}
%   \label{fig:teaser}
% \end{teaserfigure}

% \received{20 February 2007}
% \received[revised]{12 March 2009}
% \received[accepted]{5 June 2009}

%%
%% This command processes the author and affiliation and title
%% information and builds the first part of the formatted document.
\maketitle

\section{Introduction}
\begin{figure}[h]
	\centering
	\includegraphics[width=0.97\linewidth]{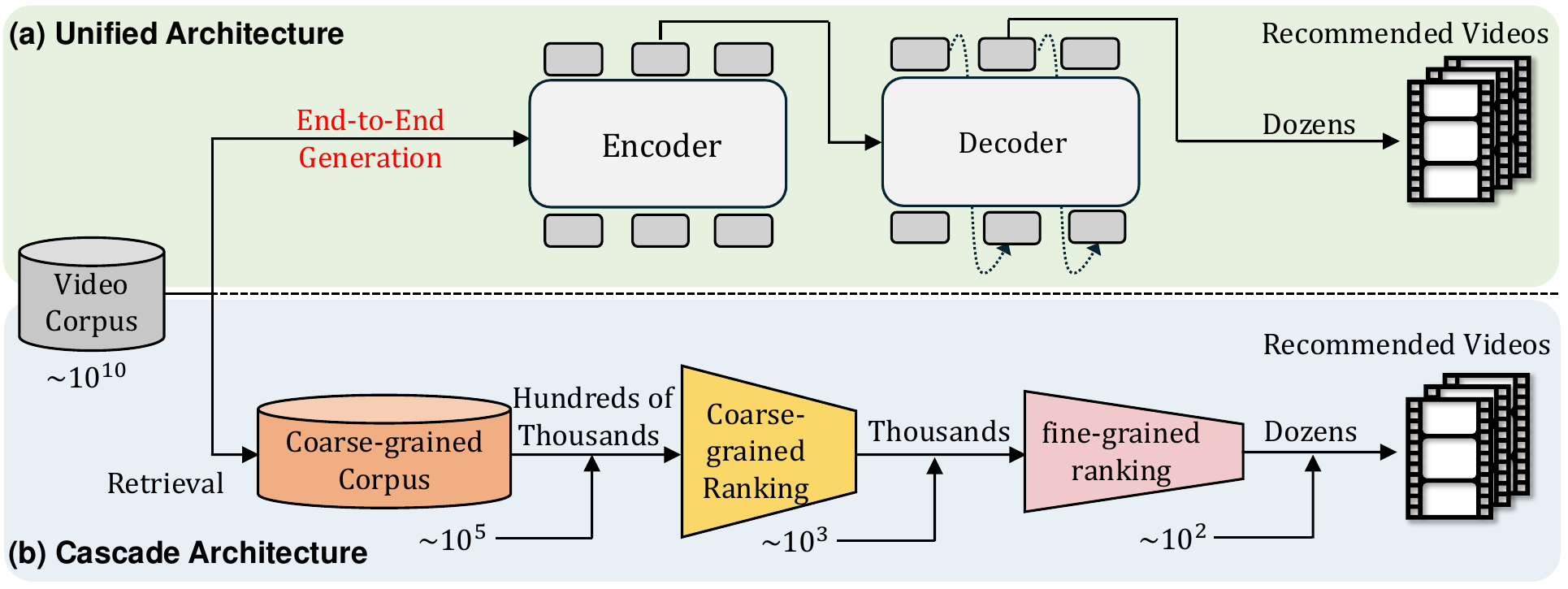}
	\vspace{-0.5mm}
	\caption{(a) Our proposed unified architecture for end-to-end generation. (b) A typical cascade ranking system, which includes three stages from the bottom to the top: Retrieval, Pre-ranking, and Ranking.}
	\label{cascadedranking}
	\vspace{-0.6cm}
    \end{figure}

    To balance efficiency and effectiveness, most modern recommender systems adopt a cascade ranking strategy\cite{covington2016deep, liu2017cascade, wang2011cascade, qin2022rankflow}. As illustrated in Figure \ref{cascadedranking}(b), a typical cascade ranking system employs a three-stage pipeline: recall \cite{covington2016deep, huang2013learning, zhu2018learning}, pre-ranking \cite{ma2021towards, wang2020cold}, and ranking \cite{burges2010ranknet, guo2017deepfm, hidasi2015session, zhou2019deep, zhou2018deep, pi2020search, chang2023twin}. Each stage is responsible for selecting the top-$k$ items from the received items and passing the results to the next stage, collectively balancing the trade-off between system response time and sorting accuracy.

    Although efficient in practice, existing methods typically treat each ranker independently, where the effectiveness of each isolated stage serves as the upper bound for the subsequent ranking stage, thereby limiting the performance of the overall ranking system. Despite various efforts \cite{gallagher2019joint, fei2021gemnn, hron2021component, qin2022rankflow, huang2023cooperative, wang2024adaptive} to improve overall recommendation performance by enabling interaction among rankers, they still maintain the traditional cascade ranking paradigm. Recently, generative retrieval-based recommendation systems (GRs) \cite{rajput2023recommender, wang2024eager, zheng2024adapting} have emerged as a promising paradigm by directly generating the identifier of a candidate item in an autoregressive sequence generation manner. By indexing items with quantized semantic IDs that encode item semantics \cite{lee2022autoregressive}, recommenders can leverage the abundant semantic information within the items. The generative nature of GRs makes them suitable for directly selecting candidate items through beam search decoding and producing more diverse recommendation results. However, current generative models only act as selectors in the retrieval stage, as their recommendation accuracy does not yet match that of well-designed multiple cascade rankers.

    To address the above challenges, we propose a unified end-to-end generative framework for single-stage recommendation named \textbf{OneRec}. \textit{First}, we present an encoder-decoder architecture. Taking inspiration from the scaling laws observed in training large language models, we find that scaling the capacity of recommendation models also consistently improves the performance. So we scale up the model parameters based on the structure of MoE \cite{zoph2022designing, du2022glam, dai2024deepseekmoe}, which significantly improves the model's ability to characterize user interests. \textit{Second}, unlike the traditional point-by-point prediction of the next item, we propose a session-wise list generation approach that considers the relative content and order of the items within each session. The point-by-point generation method necessitates hand-craft strategies to ensure coherence and diversity in the generated results. In contrast, the session-wise learning process enables the model to autonomously learn the optimal session structure  by feeding it preferred data. \textit{Last but not least}, we explore preference learning by using direct preference optimization (DPO) \cite{rafailov2024direct} to further enhance the quality of the generated results. For constructing preference pairs, we take inspiration from hard negative sampling \cite{shi2023theories} by creating self-hard rejected samples from the beam search results rather than random sampling. We propose an Iterative Preference Alignment (IPA) strategy to rank the sampled responses based on scores provided by the pre-trained reward model (RM), identifying the best-chosen and worst-rejected samples. Our experiments on large-scale industry datasets show the superiority of the proposed method. We also conduct a series of ablation experiments to demonstrate the effectiveness of each module in detail. The main contributions of this work are summarized as follows:
    \begin{itemize}[leftmargin=*]
    \item To overcome the limitations of cascade ranking, we introduce OneRec, a single-stage generative recommendation framework. 
    To the best of our knowledge, this is one of the first industrial solutions capable of handling item recommendations with a unified generation model, significantly surpassing the traditional multi-stage ranking pipeline.
    
    \item We highlight the necessity of model capacity and contextual information of target items through a session-wise generation manner, which enables more accurate predictions and enhances the diversity of generated items.
    
    \item We propose a novel self-hard negative samples selection strategy based on personalized reward model. With direct preference optimization, we enhance OneRec's generalization across a broader range of user preference. Extensive offline experiments and online A/B testing demonstrates their effectiveness and efficiency.
    \end{itemize}

% (i) the balanced identifier quantitation stage which generates the identifier for each based on pre-trained IA embedding;
\begin{figure*}[h]
\centering
\includegraphics[width=.95\textwidth]{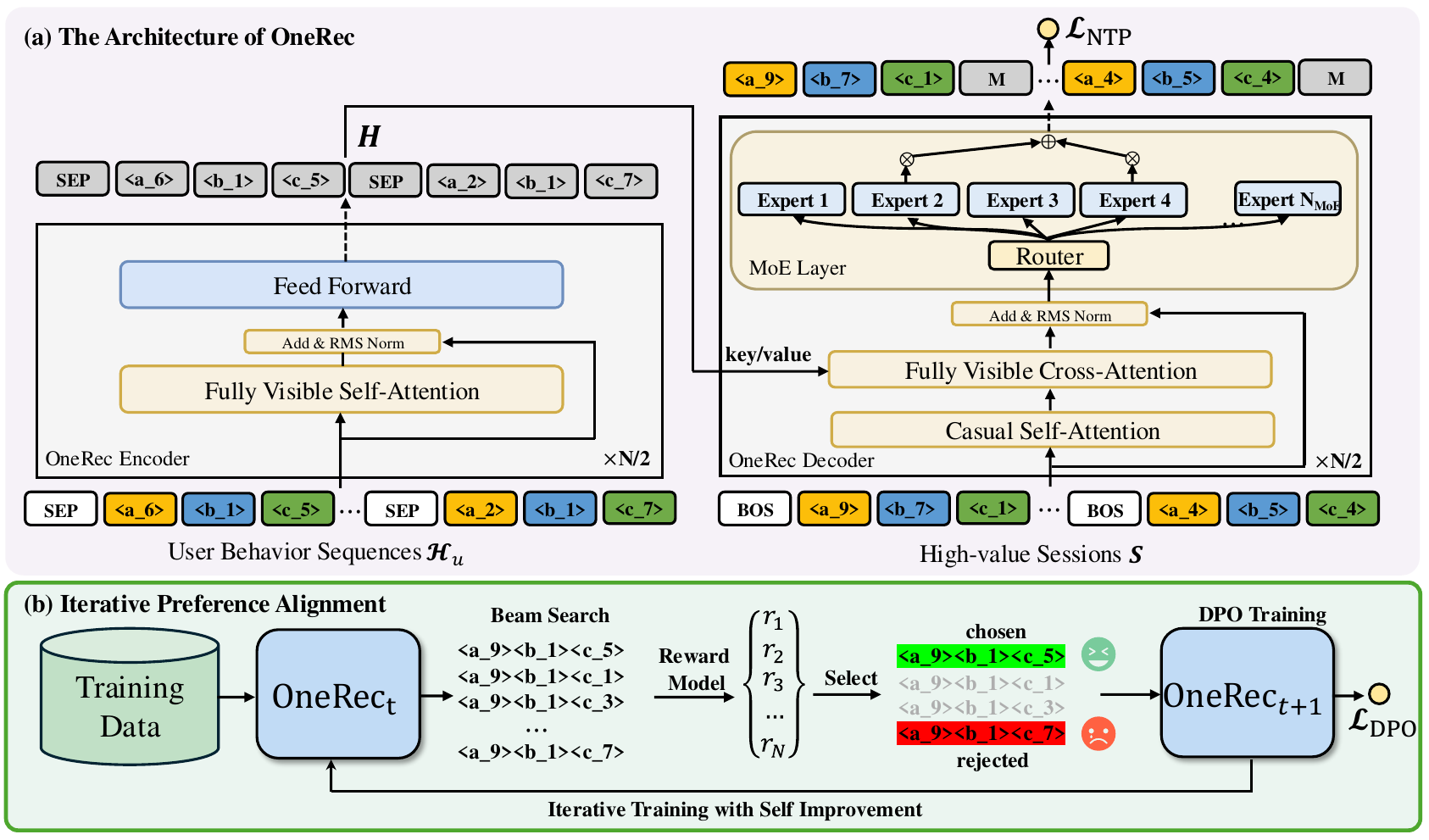}
\caption{\textbf{The overall framework of OneRec, consists of two stages: (i) the session training stage which train OneRec with session-wise data; (ii) the IPA stage which utilizes iterative direct preference optimization with self-hard negatives.}}
\label{fig2}
% \vspace{-0.4cm}
\end{figure*}

\vspace{-0.4cm}
\section{Related Work}

\subsection{Generative Recommendation}
In recent years, with the remarkable progress in generative models, generative recommendation has received increasing attention. Unlike traditional embedding-based retrieval methods which largely rely on a two-tower model for calculating the ranking score for each candidate item and utilize an effecient MIPS or ANN \cite{houle2014rank,muja2014scalable, shrivastava2014asymmetric, ge2013optimized,jegou2010product} search system for retrieving top-$k$ relevant items. Generative Retrieval (GR) \cite{tang2023recent} method formulates the problem of retrieving relevant documents from the database as a sequence generation task which generate the relevant document tokens sequentially. The document tokens can be the document titles, document IDs or pre-trained semantic IDs \cite{tay2022transformer}. GENRE \cite{de2020autoregressive} first adopts the transformer architecture for entity retrieval, generating entity names in an autoregressive fashion based on the conditioned context. DSI \cite{tay2022transformer} first proposes the concept of assigning structured semantic IDs to documents and training encoder-decoder models for generative document retrieval. Following this paradigm, TIGER \cite{rajput2023recommender} introduces the formulation of generative item retrieval models for recommender systems. 

In addition to the generation framework, how to index items has also attracted increasing attention. Recent research focuses on the semantic indexing technique \cite{rajput2023recommender,tay2022transformer, feng2022recommender}, which aims to index items based on content information. Specifically, TIGER \cite{rajput2023recommender} and LC-Rec \cite{zheng2024adapting} apply residual quantization (RQ-VAE) to textual embeddings derived from item titles and descriptions for tokenization. Recforest \cite{feng2022recommender} utilizes hierarchical k-means clustering on item textual embeddings to obtain cluster indexes as tokens. Furthermore, recent studies such as EAGER \cite{wang2024eager} explore integrating both semantic and collaborative information into the tokenization process. 

\subsection{Preference Alignment of Language Models}
During the post-training \cite{dubey2024llama} phase of Large Language Models (LLMs), Reinforcement Learning from Human Feedback (RLHF) \cite{stiennon2020learning, ouyang2022training} is a prevalent method in aligning LLMs with human values by employing reinforcement learning techniques guided by reward models that represent human feedback. However, RLHF suffers from instability and inefficiency. Direct Preference Optimization (DPO) \cite{rafailov2024direct} is proposed which derives the optimal policy in closed form and enables direct optimization using preference data. Apart from that, several variants have been proposed to further improve the original DPO. For example, IPO \cite{azar2024general} bypasses two approximations in DPO with a general objective. cDPO \cite{rafailov2024direct} alleviates the influence of noisy labels by introducing a hyperparameter $\epsilon$. rDPO \cite{chowdhury2024provably} designs an unbiased estimate of the original Binary Cross Entropy loss. Other variants including CPO \cite{xu2024contrastive}, simDPO \cite{chowdhury2024provably}, also enhance or expand
DPO in various aspects. However, unlike traditional NLP scenarios where preference data is explicitly annotated through humans, preference learning in recommendation systems faces a unique challenge because of the sparsity of user-item interaction data. This challenge results in adapting DPO for recommendation are still largely unexplored. Different from S-DPO which focuses on incorporating multiple negatives in user preference data for LM-based recommenders, we train a reward model and based on the scores from reward model we choose personalized preference data for different users.

\vspace{-0.4cm}
\section{Methods}
In this section, we propose OneRec, an end-to-end framework that generates target items through a single-stage retrieval manner. In Section \ref{biq}, we first introduce the feature engineering for the single-stage generative recommendation pipeline in industrial applications. Then, in Section \ref{srt}, we formally define the session-wise generative tasks and present the architecture of our proposed OneRec model. Finally, we elaborate on the model's capability with a personalized reward model for self-hard negative sampling in Section \ref{rpo}, and demonstrate how we iteratively improve model performance through direct preference optimization. The overall framework of OneRec is illustrated in Figure \ref{fig2}.

\vspace{-0.4cm}
\subsection{Preliminary} \label{biq} 
In this section, we introduce the construction of the single-stage generative recommendation pipeline from the perspectives of feature engineering. For user-side feature, OneRec takes the positive historical behavior sequences $\mathcal{H}_u=\{\bm{v}^{h}_1, \bm{v}^{h}_2,\ldots, \bm{v}^{h}_n\}$ as input, where $\bm{v}$ represent the videos that the user has effectively watched or interacted with (likes, follows, shares), and $n$ is the length of behaviour sequence. The output of OneRec is a list of videos, consisting of a session $\mathcal{S} = \{\bm{v}_1, \bm{v}_2, ..., \bm{v}_m\}$,  where $m$ is the number of videos within a session (the detailed definition of ``session'' can be found in Section \ref{srt}).

For each video $\bm{v}_i$, we describe them with multi-modal embeddings $\textbf{\textit{e}}_i \in \mathbb{R}^{d}$ which are aligned with the real user-item behaviour distribution \cite{luo2024qarm}. Based on the pretrain multi-modal representation, existing generative recommendation frameworks \cite{liu2024mmgrec, rajput2023recommender} use RQ-VAE \cite{zeghidour2021soundstream} to encode the embedding into semantic tokens. However, such method is suboptimal due to the unbalanced code distribution which is known as the \textit{hourglass phenomenon} \cite{kuai2024breaking}. We apply a multi-level balanced quantitative mechanism to transform the $\textbf{\textit{e}}_i$ with residual K-Means quantization algorithm\cite{luo2024qarm}. At the first level ($l = 1$), the initial residual is defined as $\bm{r}_i^{1} =\bm{e}_i$. At each level $l$, we have a codebook $\mathcal{C}_{l}=\{\bm{c}_1^{l},...,\bm{c}_K^{l}\} $, where $K$ is the codebook size. The index of the closest centroid node embedding is generate through $\bm{s}_i^l = \arg\min_{k}\|\textbf{\textit{r}}_i^{l} - \bm{c}_k^{l}\|_2^2$ and for next level $l + 1$ the residual is defined as $\bm{r}_i^{l + 1} =\bm{r}_i^{l} -  \bm{c}_{\bm{s}_i^l}^{l}  $. So the corresponding codebook tokens are generated through hierarchical indexing:
\[
\begin{aligned}
&\bm{s}_i^1 = \arg\min_{k}\|\bm{r}_i^{1} - \bm{c}_k^{1}\|_2^2, \quad \bm{r}_i^{2} =\bm{r}_i^{1} -  \bm{c}_{\bm{s}_i^1}^{1}  \\
&\bm{s}_i^2 = \arg\min_{k}\|\bm{r}_i^{2} - \bm{c}_k^{2}\|_2^2, \quad \bm{r}_i^{3} =\bm{r}_i^{2} -  \bm{c}_{\bm{s}_i^2}^{2}  \\
&\qquad\qquad\vdots \\
&\bm{s}_i^L = \arg\min_{k}\|\bm{r}_i^{L} - \bm{c}_k^{L}\|_2^2 \\
\end{aligned}
\]
where $L$ is the total layers of sematic ID.

To construct a balanced codebook $\mathcal{C}_l = \{\bm{c}_1^l, \ldots, \bm{c}_K^l\}$, we apply the Balanced K-means as detailed in \textbf{Algorithm \ref{kmeans}} for itemset partitioning. Given the total video set $\mathcal{V}$, this algorithm partitions the set into $K$ clusters, where each cluster contains exactly $w = |\mathcal{V}|/K$ videos. During iterative computation, each centroid is sequentially assigned its $w$ nearest unallocated videos based on Euclidean distance, followed by centroid recalibration using mean vectors of assigned videos. The termination criterion is satisfied when cluster assignments reach convergence.

\begin{algorithm}[t]
  \SetAlgoLined
  \caption{Balanced K-means Clustering}
  \label{kmeans}

  \KwIn{Item set $\mathcal{V}$, number of clusters $K$}

  Compute $w \gets |\mathcal{V}|/K$

  Initialize centroids $\mathcal{C}_l = \{\bm{c}_1^l, \ldots, \bm{c}_K^l\}$ with random selection\;

  \Repeat{Assignment convergence}{
  Initialize unassigned set $\mathcal{U} \gets \mathcal{V}$

  \For{each cluster $k \in \{1, \ldots, K\}$}{
      Sort $\mathcal{U}$ by ascending distance from centroid $\bm{c}_k^l$\;

      Assign $\mathcal{V}_k \gets \mathcal{U}[0:w-1]$\;

      Update centroid $\bm{c}_k^l \gets \frac{1}{w}\sum_{\bm{r}^l \in \mathcal{V}_k}\bm{r}^l$\;

      Remove assigned items $\mathcal{U} \gets \mathcal{U} \setminus \mathcal{V}_k$\;
    }
  }
  \KwOut{Optimized codebook $\mathcal{C}_l = \{\bm{c}_1^l, \ldots, \bm{c}_K^l\}$}
  
\end{algorithm}

\subsection{Session-wise List Generation} \label{srt}
Different from traditional point-wise recommendation methods that only predict the next video, session-wise generation aims to generate a list of high-value sessions based on users' historical interaction sequences, which enables the recommendation model to capture the dependencies between videos in the recommended list. Specifically, a session refers to a batch of short videos returned in response to a user's request, typically consisting of 5 to 10 videos. The videos within a session generally take into account factors such as user interest, coherence, and diversity. We have devised several criteria to identify high-quality sessions, including:
\begin{itemize}[leftmargin=*]
\item The number of short videos actually watched by the user within a session is greater than or equal to 5;
\item The total duration for which the user watches the session exceeds a certain threshold;
\item The user exhibits interaction behaviors, such as liking, collecting, or sharing the videos;

\end{itemize}

This approach ensures that our session-wise model learns from real user engagement patterns and captures more accurate contextual information within the session list. So the objective of our session-wise model $\mathcal{M}$ can be formalized as: 
\begin{equation}
\mathcal{S} :=\mathcal{M}(\mathcal{H}_u)
\end{equation}
where $\mathcal{H}_u$ is represented from the remantic IDs: $\mathcal{H}_u = \{ (\bm{s}_1^1, \bm{s}_1^2, \cdots, \\ \bm{s}_1^L),  (\bm{s}_2^1, \bm{s}_2^2, \cdots, \bm{s}_2^L), \cdots, (\bm{s}_n^1, \bm{s}_n^2, \cdots, \bm{s}_n^L) \}$ and $\mathcal{S} = \{ (\bm{s}_1^1, \bm{s}_1^2, \cdots, \bm{s}_1^L),\\ (\bm{s}_2^1, \bm{s}_2^2, \cdots, \bm{s}_2^L), \cdots, (\bm{s}_m^1, \bm{s}_m^2, \cdots, \bm{s}_m^L) \}$.

As shown in Figure \ref{fig2} (a), consistent with the T5 \cite{xu2024openp5} architecture, our model employs a transformer-based framework consisting of two main components: an encoder for modeling user historical interactions and a decoder for session list generation. Specifically, the encoder leverages the stacked multi-head self-attention and feed-forward layers to process the input sequence $\mathcal{H}_u$. We denote the encoded historical interaction features as $\textbf{\textit{H}}=Encoder(\mathcal{H}_u)$.

The decoder takes the semantic IDs of the target session as input and generates the target in an auto-regressive manner. To train a larger model at reasonable economic costs, for the feed-forward neural networks (FNNs) in the decoder, we adopt the MoE architecture \cite{zoph2022designing, du2022glam, dai2024deepseekmoe} commonly used in Transformer-based language models and substitute the $l$-th FFN with: 
\begin{equation}
\begin{split}
\mathbf{H}_{t}^{l+1} & = \sum_{i=1}^{N_{\rm MoE}} \left( {g_{i,t} \operatorname{FFN}_{i}\left( \mathbf{H}_{t}^{l} \right)} \right) + \mathbf{H}_{t}^{l}, \\
g_{i,t} & = \begin{cases} 
s_{i,t}, & s_{i,t} \in \operatorname{Topk} (\{ s_{j, t} | 1 \leq j \leq N \}, K_{\rm MoE}), \\
0, & \text{otherwise}, 
\end{cases} \\
s_{i,t} & = \operatorname{Softmax}_i \left( {\mathbf{H}_{t}^{l}}^{T} \mathbf{e}_{i}^{l} \right), 
\end{split}
\end{equation}
where $N_{\rm MoE}$ represents the total number of experts, $\operatorname{FFN}_{i}(\cdot)$ is the $i$-th expert FFN, and $g_{i,t}$ denotes the gate value for the $i$-th expert. The gate value $g_{i,t}$ is sparse, meaning that only $K_{\rm MoE}$ out of $N_{\rm MoE}$ gate values are non-zero. This sparsity property ensures computational efficiency within an MoE layer and each token will be assigned to and computed in only $K_{\rm MoE}$ experts. 

For training, we add a start token $\bm{s}_{[\rm BOS]}$ at the beginning of codes to construct the decoder inputs with:
\begin{equation}
\begin{gathered}
{\mathcal{\bar S}}=\{\bm{s}_{[\rm BOS]},\bm{s}_1^1, \bm{s}_1^2, \cdots, \bm{s}_1^L, \bm{s}_{[\rm BOS]}, \bm{s}_2^1, \bm{s}_2^2,\cdots, \bm{s}_2^L,
\\
\cdots, \bm{s}_{[\rm BOS]},\bm{s}_m^1, \bm{s}_m^2, \cdots, \bm{s}_m^L \}
\end{gathered}
\end{equation}

We utilize cross-entropy loss for next-token prediction on the sematic IDs of the target session. The NTP loss $\mathcal{L}_{\rm NTP}$ is formulated as:
\begin{equation}
\begin{gathered}
\mathcal{L}_{\rm NTP} = - \sum_{i=1}^m \sum_{j=1}^L \log P( \bm{s}_{i}^{j+1} \mid [\bm{s}_{[\rm BOS]},\bm{s}_1^1, \bm{s}_1^2, \cdots,  \bm{s}_1^L, \cdots,
\\
\bm{s}_{[\rm BOS]},\bm{s}_i^1,  \cdots,  \bm{s}_i^j ]; \Theta).
\end{gathered}
\end{equation}

After a certain amount of training on session-wise list generation task, we obtain the seed model $\mathcal{M}_t$.

\subsection{Iterative Preference Alignment with RM} \label{rpo}
The high-quality sessions defined in Section \ref{srt} provide valuable training data, enabling the model to learn what constitutes a good session, thereby ensuring the quality of generated videos. Building on this foundation, we aim to further enhance the model's ability by Direct Preference Optimization (DPO). In traditional natural language processing (NLP) scenarios, preference data is explicitly annotated by humans. However, preference learning in recommendation systems confronts a unique challenge due to the sparsity of user-item interaction data, which necessitates a reward model (RM). So we introduce a session-wise reward model in Section \ref{rewardmodel}. Moreover, we improve the conventional DPO by proposing an iterative direct preference optimization that enables the model to self-improvement described in Section \ref{Iterative}.

\begin{algorithm}[t]
  \SetAlgoLined
  \caption{Iterative Preference Alignment (IPA)}
  \label{rpoalg}
  
  \KwIn{Number of responses $N$, pretrained RM $R(\bm{u},\mathcal{S})$,
        seed model $\mathcal{M}_t$, DPO ratio $r_{\mathrm{DPO}}$,
        total epochs $T$ and samples per epoch $N_{\mathrm{sample}}$}
  
  \For{$\mathit{epoch} \gets t$ \KwTo $T$}{  % 规范epoch计数从1开始
    \For{$\mathit{sample} \gets 1$ \KwTo $N_{\mathrm{sample}}$}{
      \eIf{$\mathit{rand}() < r_{\mathrm{DPO}}$}{  % 概率判断更规范的写法
        Generate $N$ responses via $\mathcal{M}_t$:
        
        \For{$i \gets 1$ \KwTo $N$}{
          $\mathcal{S}_u^i \sim \mathcal{M}_t(\mathcal{H}_u)$\; % 补全模型符号
          $r_u^i \gets R(\bm{u}, \mathcal{S}_u^i)$\; % 合并评估步骤
        }
        Select the best and worst responses:
        
        $\mathcal{S}_u^w \gets \mathcal{S}_u^{\arg\max_i r_u^i}$\; % 明确索引关系
        $\mathcal{S}_u^l \gets \mathcal{S}_u^{\arg\min_i r_u^i}$\;
        
        Compute NTP and DPO loss:
        
        $\mathcal{L} \gets \mathcal{L}_{\mathrm{NTP}} + \lambda\mathcal{L}_{\mathrm{DPO}}$\;
      }{
        Compute NTP loss:
        
        $\mathcal{L} \gets \mathcal{L}_{\mathrm{NTP}}$\;
      }
      Update parameters:
      
      $\Theta \gets \Theta - \alpha\nabla_\Theta\mathcal{L}$\; % 改用梯度符号
    }
    Update model snapshot: $\mathcal{M}_{t+1} \gets \mathcal{M}_t$\;
  }
  \KwOut{Optimized parameters $\Theta$}
\end{algorithm}
\subsubsection{Reward Model Training}
\label{rewardmodel}

We use $R(\bm{u},\mathcal{S})$ to denote the reward model which selects preference data for different users. Here, the output $r$ represents the reward corresponding to user $u$'s (usually represented by user behavior) preference on the session $\mathcal{S}=\{\bm{v}_1, \bm{v}_2,\ldots, \bm{v}_m\}$. 
In order to equip the RM with the capacity to rank session, we first extract the target-aware representation $\bm{e}_i =  \bm{v}_{i} \odot \bm{u}$ of each item $\bm{v}_i$ in $\mathcal{S}$, where $\odot$ represents the target-aware operation (such as target attention toward user behavior). So we get the target-aware representation $\boldsymbol{h} = \{\bm{e}_1, \bm{e}_2 , \cdots, \bm{e}_m \}$ for session $\mathcal{S}$. Then the items within a session interact with each other through self-attention layers to fuse the necessary information among different items:
\begin{small}
\begin{equation}
\begin{aligned}
\boldsymbol{h}_f = \mathrm{SelfAttention}(\boldsymbol{h} \boldsymbol{W}^Q_s, \boldsymbol{h}\ \boldsymbol{W}^K_s, \boldsymbol{h} \boldsymbol{W}^V_s)
\end{aligned}
\end{equation}
\end{small}

Next we utilize different tower to make predictions on multi-target reward and the RM is pre-trained with abundant recommendation data:
\begin{equation}
\small
% \footnotesize
\begin{split}
\hat{r}^{swt} &= \texttt{Tower}^{swt}\big(\texttt{Sum}\big( \boldsymbol{h}_f  \big)\big), 
\hat{r}^{vtr} = \texttt{Tower}^{vtr}\big(\texttt{Sum}\big( \boldsymbol{h}_f  \big)\big),\\
\hat{r}^{wtr} &= \texttt{Tower}^{wtr}\big(\texttt{Sum}\big( \boldsymbol{h}_f  \big)\big), 
\hat{r}^{ltr} = \texttt{Tower}^{ltr}\big(\texttt{Sum}\big( \boldsymbol{h}_f  \big)\big), \\
&\texttt{whe}\texttt{re}\quad\texttt{Tower}(\cdot) = \texttt{Sigmoid}\big(\texttt{MLP}(\cdot)\big)
\end{split}
\label{cgc}
\end{equation}

After getting all the estimated rewards $\hat{r}^{swt}, \dots$ and the ground-truth labels $y^{swt}, \dots$ for each session, we directly minimize the binary cross-entropy loss to train the RM. The loss function $\mathcal{L}_{\rm RM}$ is defined as follows:
\begin{equation}
\begin{split}
\mathcal{L}_{\rm RM} =-\sum_{{swt,\dots}}^{xtr}\left(y^{xtr}\log{(\hat{r}^{xtr})}+(1 - y^{xtr})\log{(1-\hat{r}^{xtr})}\right)
\end{split}\label{crossentropy}
\end{equation}

\subsubsection{Iterative Preference Alignment}
\label{Iterative}
Based on pre-trained RM $R(\bm{u},\mathcal{S})$ and current OneRec $\mathcal{M}_t$, we generate $N$ different
responses for each user by beam search:
\begin{equation}
\mathcal{S}_u^n \sim M_t(\mathcal{H}_u) \quad \text{for all} \  u \in \mathcal{U} \  \text{and} \  n \in [N],
\end{equation}
where we use $[N]$ to denote $\{1,2,\dots,N\}$.

Then we computes the reward $r_u^n$ for each of these responses based on the RM $R(\bm{u},\mathcal{S})$:
\begin{equation}
r_u^n=R(\bm{u},\mathcal{S}_u^n)
\end{equation}

Next we build the preference pairs $D_t^\text{pairs} = (\mathcal{S}_u^w,\mathcal{S}_u^l,\mathcal{H}_u)$ by choosing the winner response $(\mathcal{S}_u^w,\mathcal{H}_u)$ with the highest reward value and a loser response $(\mathcal{S}_u^l,\mathcal{H}_u)$ with the lowest reward value. Given the preference pairs, we can now train a new model $M_{t + 1}$ which is initialized from model $M_{t}$, and updated with a loss function that combines the DPO loss \citep{rafailov2024direct} for learning from the preference pairs. The loss corresponding to each preference pair is as follows:
\begin{equation}
\begin{split}    
    \mathcal{L}_\text{DPO} &=  \mathcal{L}_\text{DPO}( \mathcal{S}_u^w, \mathcal{S}_u^l | \mathcal{H}_u) \\
    &=  - \log \sigma \left( \beta \log \frac{ M_{t+1}( \mathcal{S}_u^w | \mathcal{H}_u)}{ M_{t}(\mathcal{S}_u^w  | \mathcal{H}_u)} - \beta \log \frac{ M_{t+1}(\mathcal{S}_u^l| \mathcal{H}_u)}{ M_{t}(\mathcal{S}_u^l| \mathcal{H}_u)}\right).
\label{eq:loss}
\end{split}
\end{equation}

As shown in Algorithm \ref{rpoalg} and Figure \ref{fig2} (b), the overall procedure involves training a sequence of models $M_t,\dots,M_T$. To mitigate the computational burden during beam search inference, we randomly sample only $r_{\rm DPO} = 1\%$ data for preference alignment. For each successive model $M_{t + 1}$, it initializes from previous model $M_{t}$ and utilizes the preference data $D_t^\text{pairs}$ generated by the $M_{t}$ for training.

\begin{table*}[!t]
\centering
\caption{
Offline performance of our proposed \emphcolorgreenn{OneRec (green)} with \emphcolorbrownn{pointwise methods (brown)}, \emphcolorbluee{listwise methods (blue)} and \emphcoloryelloww{preference alignment methods (yellow)}. Best results are in bold, sub-optimal results are underlined. Metrics with $\uparrow$ indicate higher is better, while $\downarrow$ indicates lower is better.
}
\label{overall}
\tabcolsep=2pt
\renewcommand\arraystretch{0.9}
\resizebox{0.99\linewidth}{!}{
\begin{tabular}{
    >{\raggedright\arraybackslash}p{2.6cm}  % 第一列左对齐
    >{\centering\arraybackslash}p{1.8cm}    % 调整后共9列
    >{\centering\arraybackslash}p{1.8cm}
    >{\centering\arraybackslash}p{1.8cm}
    >{\centering\arraybackslash}p{1.8cm}
    >{\centering\arraybackslash}p{2.0cm}
    >{\centering\arraybackslash}p{2.0cm}
    >{\centering\arraybackslash}p{1.8cm}
    >{\centering\arraybackslash}p{1.8cm}
    % >{\centering\arraybackslash}p{1.8cm}
}
\toprule
\multirow{3}{*}{\textbf{Model}} & 
\multicolumn{4}{c}{\textit{Watching-Time Metrics}} &
\multicolumn{4}{c}{\textit{Interaction Metrics}} \\  % 删除Diversity跨列
& \multicolumn{2}{c}{\textbf{swt$\uparrow$}} & \multicolumn{2}{c}{\textbf{vtr$\uparrow$}}
& \multicolumn{2}{c}{\textbf{wtr$\uparrow$}} & \multicolumn{2}{c}{\textbf{ltr$\uparrow$}} \\
& mean & max & mean & max & mean & max & mean & max \\ \midrule  % 删除UCN/ILS行

\multicolumn{9}{c}{\textit{Pointwise Discriminative Method}} \\  \cdashlinelr{1-9}  % 调整列范围
\cellcolor{brown!15}SASRec &
  \cellcolor{brown!15} 0.0375$\pm$0.002&
  \cellcolor{brown!15} 0.0803$\pm$0.005&
  \cellcolor{brown!15} 0.4313$\pm$0.013&
  \cellcolor{brown!15} 0.5801$\pm$0.013&
  \cellcolor{brown!15} 0.00294$\pm$0.001&
  \cellcolor{brown!15} \underline{0.00978$\pm$0.001}&
  \cellcolor{brown!15} 0.0314$\pm$0.002&
  \cellcolor{brown!15} 0.0604$\pm$0.004\\
\cellcolor{brown!15}BERT4Rec &
  \cellcolor{brown!15} 0.0336$\pm$0.002&
  \cellcolor{brown!15} 0.0706$\pm$0.004&
  \cellcolor{brown!15} 0.4192$\pm$0.014&
  \cellcolor{brown!15} 0.5633$\pm$0.013&
  \cellcolor{brown!15} 0.00281$\pm$0.001&
  \cellcolor{brown!15} 0.00932$\pm$0.001&
  \cellcolor{brown!15} 0.0316$\pm$0.002&
  \cellcolor{brown!15} 0.0606$\pm$0.004\\
\cellcolor{brown!15}FDSA &
  \cellcolor{brown!15} 0.0325$\pm$0.002&
  \cellcolor{brown!15} 0.0683$\pm$0.005&
  \cellcolor{brown!15} 0.4145$\pm$0.015&
  \cellcolor{brown!15} 0.5588$\pm$0.014&
  \cellcolor{brown!15} 0.00271$\pm$0.001&
  \cellcolor{brown!15}0.00921$\pm$0.001&
  \cellcolor{brown!15}0.0313$\pm$0.002&
  \cellcolor{brown!15} 0.0604$\pm$0.003\\  \cdashlinelr{1-9}

\multicolumn{9}{c}{\textit{Pointwise Generative Method}} \\  \cdashlinelr{1-9}
  \cellcolor{brown!15}TIGER-0.1B &
  \cellcolor{brown!15} 0.0879$\pm$0.007&
  \cellcolor{brown!15} 0.1286$\pm$0.010&
  \cellcolor{brown!15} 0.5826$\pm$0.016&
  \cellcolor{brown!15} 0.6625$\pm$0.017&
  \cellcolor{brown!15} 0.00277$\pm$0.001&
  \cellcolor{brown!15} 0.00671$\pm$0.001&
  \cellcolor{brown!15} 0.0316$\pm$0.004&
  \cellcolor{brown!15} 0.0541$\pm$0.007 \\
\cellcolor{brown!15}TIGER-1B &
\cellcolor{brown!15} 0.0873$\pm$0.006&
\cellcolor{brown!15} 0.1368$\pm$0.010&
\cellcolor{brown!15} 0.5827$\pm$0.015&
\cellcolor{brown!15} 0.6776$\pm$0.015&
\cellcolor{brown!15} 0.00292$\pm$0.001&
\cellcolor{brown!15} 0.00758$\pm$0.001&
\cellcolor{brown!15} 0.0323$\pm$0.004&
\cellcolor{brown!15} 0.0579$\pm$0.008 \\ \cdashlinelr{1-9}

\multicolumn{9}{c}{\textit{Listwise Generative Method}} \\  \cdashlinelr{1-9}
  \cellcolor{gblue!10}OneRec-0.1B&
  \cellcolor{gblue!10}0.0973$\pm$0.010&
  \cellcolor{gblue!10}0.1501$\pm$0.015&
  \cellcolor{gblue!10} 0.6001$\pm$0.021&
  \cellcolor{gblue!10} 0.6981$\pm$0.021&
  \cellcolor{gblue!10}0.00326$\pm$0.001&
  \cellcolor{gblue!10}0.00870$\pm$0.001&
  \cellcolor{gblue!10}0.0349$\pm$0.009&
  \cellcolor{gblue!10}0.0642$\pm$0.015 \\
  
  \cellcolor{gblue!10}OneRec-1B &
  \cellcolor{gblue!10}0.0991$\pm$0.008&
  \cellcolor{gblue!10}0.1529$\pm$0.012&
  \cellcolor{gblue!10} 0.6039$\pm$0.020&
  \cellcolor{gblue!10} 0.7013$\pm$0.020&
  \cellcolor{gblue!10} \underline{0.00349$\pm$0.001}&
  \cellcolor{gblue!10} 0.00919$\pm$0.002&
  \cellcolor{gblue!10} 0.0360$\pm$0.005&
  \cellcolor{gblue!10} \underline{0.0660$\pm$0.008} \\  \cdashlinelr{1-9}

\multicolumn{9}{c}{\textit{Listwise Generative Method with Preference Alignment}} \\  \cdashlinelr{1-9}
  \cellcolor{yellow!8}OneRec-1B$+$DPO &
  \cellcolor{yellow!8} 0.1014$\pm$0.010&
  \cellcolor{yellow!8} \underline{0.1595$\pm$0.015}&
  \cellcolor{yellow!8} \underline{0.6127$\pm$0.017}&
  \cellcolor{yellow!8} \underline{0.7116$\pm$0.016}&
  \cellcolor{yellow!8} 0.00339$\pm$0.001&
  \cellcolor{yellow!8} 0.00896$\pm$0.001&
  \cellcolor{yellow!8} 0.0351$\pm$0.004&
  \cellcolor{yellow!8} 0.0644$\pm$0.008 \\

  \cellcolor{yellow!8}OneRec-1B$+$IPO &
  \cellcolor{yellow!8} 0.0979$\pm$0.003&
  \cellcolor{yellow!8} 0.1528$\pm$0.005&
  \cellcolor{yellow!8} 0.6000$\pm$0.007&
  \cellcolor{yellow!8} 0.7012$\pm$0.007&
  \cellcolor{yellow!8} 0.00335$\pm$0.001&
  \cellcolor{yellow!8} 0.00905$\pm$0.001&
  \cellcolor{yellow!8} 0.0350$\pm$0.003&
  \cellcolor{yellow!8} 0.0654$\pm$0.004 \\

  \cellcolor{yellow!8}OneRec-1B$+$cDPO &
  \cellcolor{yellow!8} 0.0993$\pm$0.006&
  \cellcolor{yellow!8} 0.1547$\pm$0.008&
  \cellcolor{yellow!8} 0.6030$\pm$0.011&
  \cellcolor{yellow!8} 0.7030$\pm$0.009&
  \cellcolor{yellow!8} 0.00339$\pm$0.001&
  \cellcolor{yellow!8} 0.00901$\pm$0.001&
  \cellcolor{yellow!8} 0.0355$\pm$0.006&
  \cellcolor{yellow!8} 0.0652$\pm$0.009 \\

  \cellcolor{yellow!8}OneRec-1B$+$rDPO &
  \cellcolor{yellow!8} 0.1005$\pm$0.006&
  \cellcolor{yellow!8} 0.1555$\pm$0.008&
  \cellcolor{yellow!8} 0.6071$\pm$0.014&
  \cellcolor{yellow!8} 0.7059$\pm$0.011&
  \cellcolor{yellow!8} 0.00339$\pm$0.001&
  \cellcolor{yellow!8} 0.00899$\pm$0.001&
  \cellcolor{yellow!8} 0.0357$\pm$0.004&
  \cellcolor{yellow!8} 0.0657$\pm$0.006 \\

  \cellcolor{yellow!8}OneRec-1B$+$CPO &
  \cellcolor{yellow!8} 0.0993$\pm$0.008 &
  \cellcolor{yellow!8} 0.1538$\pm$0.012&
  \cellcolor{yellow!8} 0.6045$\pm$0.021&
  \cellcolor{yellow!8} 0.7029$\pm$0.018&
  \cellcolor{yellow!8} 0.00343$\pm$0.001&
  \cellcolor{yellow!8} 0.00911$\pm$0.002&
  \cellcolor{yellow!8} 0.0357$\pm$0.008&
  \cellcolor{yellow!8} 0.0659$\pm$0.014\\

  \cellcolor{yellow!8}OneRec-1B$+$simPO &
  \cellcolor{yellow!8} 0.0995$\pm$0.008&
  \cellcolor{yellow!8} 0.1536$\pm$0.013&
  \cellcolor{yellow!8} 0.6047$\pm$0.016&
  \cellcolor{yellow!8} 0.7022$\pm$0.015&
  \cellcolor{yellow!8} \underline{0.00349$\pm$0.001}&
  \cellcolor{yellow!8} 0.00918$\pm$0.001&
  \cellcolor{yellow!8} 0.0360$\pm$0.005&
  \cellcolor{yellow!8} 0.0659$\pm$0.008 \\

  \cellcolor{yellow!8}OneRec-1B$+$S-DPO &
  \cellcolor{yellow!8} \underline{0.1021$\pm$0.008}&
  \cellcolor{yellow!8} 0.1575$\pm$0.013&
  \cellcolor{yellow!8} 0.6096$\pm$0.016&
  \cellcolor{yellow!8} 0.7070$\pm$0.015&
  \cellcolor{yellow!8} 0.00345$\pm$0.001&
  \cellcolor{yellow!8} 0.00909$\pm$0.001&
  \cellcolor{yellow!8} \underline{0.0361$\pm$0.004}&
  \cellcolor{yellow!8} 0.0659$\pm$0.008 \\ \cdashlinelr{1-9}

  \cellcolor{green!8}OneRec-1B$+$IPA &
  \cellcolor{green!8} \textbf{0.1025$\pm$0.009} &
  \cellcolor{green!8} \textbf{0.1933$\pm$0.017}&
  \cellcolor{green!8} \textbf{0.6141$\pm$0.020}&
  \cellcolor{green!8} \textbf{0.7646$\pm$0.021}&
  \cellcolor{green!8}\textbf{0.00354$\pm$0.001}&
  \cellcolor{green!8}\textbf{0.00992$\pm$0.001}&
  \cellcolor{green!8} \textbf{0.0397$\pm$0.004}&
  \cellcolor{green!8} \textbf{0.1203$\pm$0.010} \\
  \bottomrule
\end{tabular}%
}
\label{table1}
\end{table*}

\begin{figure}[h]
\centering
\includegraphics[width=0.99\linewidth]{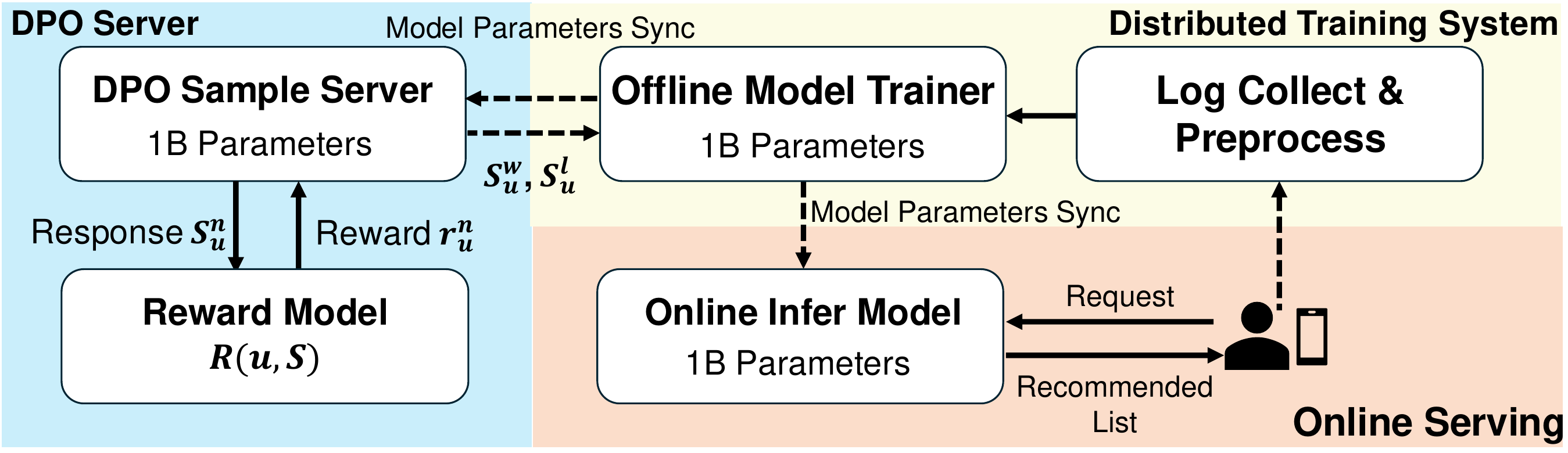}
\caption{Framework of Online Deployment of OneRec.}
\label{online}
\end{figure}

% \vspace{-0.6cm}
\section{System Deployment}
OneRec has been successfully implemented in real-world industrial scenarios. Balancing stability and performance, we deploy the OneRec-1B for online services. As illustrated in Figure \ref{online}, our deployment architecture consists of three core components: 1) the training system, 2) the online serving system, and 3) the DPO sample server. The system processes collected interaction logs as training data, initially adopting the next token prediction objective $\mathcal{L}_{\rm NTP}$ to train the seed model. After convergence, we add the DPO loss $\mathcal{L}_{\rm DPO}$ for preference alignment, leveraging XLA and bfloat16 mixed-precision training to optimize computational efficiency and memory utilization. The trained parameters are synchronized to the online inference module and the DPO sampling server for real-time serving and preference-based data selection. To enhance inference performance, we implement two key optimizations: the key-value cache decoding mechanism combined with float16 quantization to reduce GPU memory overhead, and the beam search configuration with beam size of 128 to balance generation quality and latency. Additionally, thanks to the MoE architecture, during inference only 13\% of the parameters are activated.

% \vspace{-0.4cm}
\section{Experiment}
In this section, we first compare OneRec with the point-wise methods and several DPO variations in offline settings. Then, we conduct some ablation experiments on our proposed module to verify the effectiveness of OneRec. Finally, we deploy OneRec to the online and conduct A/B test to further validate its performance on Kuaishou. 

\subsection{Experimental Settings}
\subsubsection{Implementation Details}
Our model is trained using the Adam optimizer with an initial learning rate of $2 \times 10^{-4}$. We utilize NVIDIA A800 GPUs for OneRec optimization. The DPO sample ratio $r_{\text{DPO}}$ is set to 1\% throughout training and  we generate $N = 128$ different responses for each user by beam search; The semantic identifier clustering process employs $K=8192$ clusters for each codebook layer and the number of codebook layers is set to $L=3$; The Mixture-of-Experts architecture contains $N_{\text{MoE}}=24$ expert with $K_{\text{MoE}}=2$ experts activated per forward pass through top-$k$ selection; For session modeling, we consider $m=5$ target session items and adopt $n=256$ historical behavior as context.

\subsubsection{Baseline Methods}
We adopt the following representative recommendation models, DPO and its variants to serve as additional baselines for comparison. The baseline methods include:
\begin{itemize}[leftmargin=*]
\item \textbf{SASRec} \cite{kang2018self} employs a unidirectional Transformer architecture to capture sequential dependencies in user-item interactions for next-item prediction.

\item \textbf{BERT4Rec} \cite{sun2019bert4rec} leverages bidirectional Transformers with masked language modeling to learn contextual item representations through sequence reconstruction.

\item \textbf{FDSA} \cite{zhang2019feature} implements dual self-attention pathways to jointly model item-level transitions and feature-level transformation patterns in heterogeneous recommendation scenarios.

\item \textbf{TIGER} \cite{rajput2023recommender} leverages hierarchical semantic identifiers and generative retrieval techniques for sequential recommendation through auto-regressive sequence generation.

\item \textbf{DPO} \cite{rafailov2024direct} formalizes preference optimization with a closed-form reward function derived from human feedback data via implicit reward modeling.

\item \textbf{IPO} \cite{azar2024general} proposes a theoretically grounded preference optimization framework which bypass the approximations inherent in standard DPO.

\item \textbf{cDPO} \cite{mitchellnote} introduces a robustness-aware variant incorporating a label flipping rate parameter $\epsilon $ to account for noisy preference annotations.

\item \textbf{rDPO} \cite{chowdhury2024provably} develops an unbiased loss estimator using importance sampling to reduce variance in preference optimization.

\item \textbf{CPO} \cite{xu2024contrastive} unifies contrastive learning with preference optimization through joint training of sequence likelihood rewards and supervised fine-tuning objectives.

\item \textbf{simPO} \cite{meng2024simpo} conducts preference optimization by employing sequence-level reward margins while eliminating reference model dependencies through normalized probability averaging.

\item \textbf{S-DPO} \cite{chen2024on} adapts DPO for recommendation systems through hard negative sampling and multi-item contrastive learning to enhance ranking accuracy.
\end{itemize}
\subsubsection{Evaluation Metric} We evaluate the model's performance with several key metrics. Each metric serves a distinct purpose in assessing different aspects of the model's output and we conduct the evaluation on a randomly sampled set of test cases in each iteration. To estimate the probabilities of various interactions for each specific user-session pair, we employ the pre-trained reward model to assess the value of recommended sessions. We calculate the mean reward for different target metrics, including session watch time (\textbf{swt}), view probability (\textbf{vtr}), follow probability (\textbf{wtr}) and like probability (\textbf{ltr}). Among these targets, swt and vtr are watching-time metrics, while wtr and ltr are interaction metrics.

\begin{figure}[h]
\centering
\includegraphics[width=.49\textwidth]{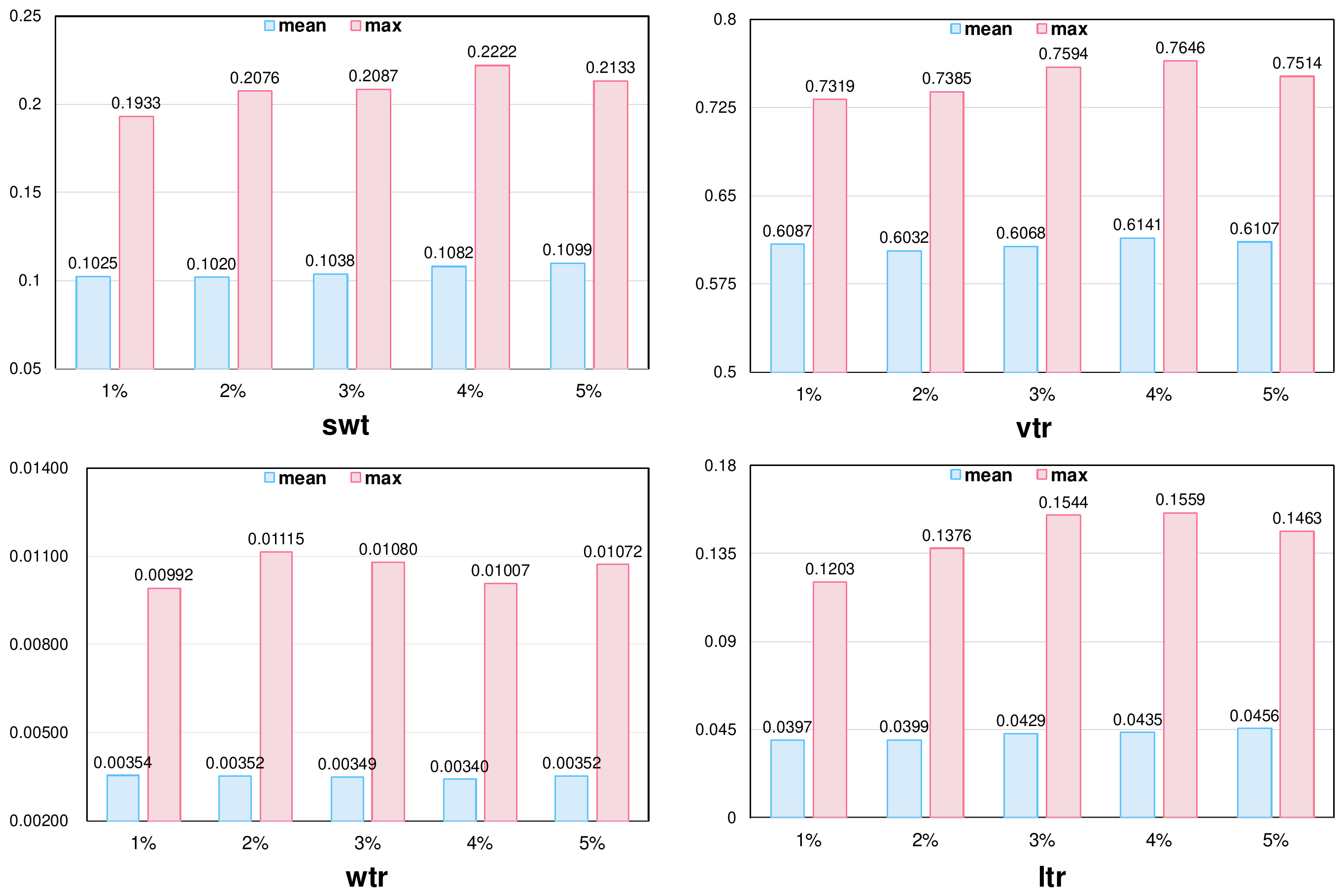}
\caption{\textbf{The ablation study on DPO sample ratio $r_{\rm DPO}$. The results indicate that a 1\% ratio of DPO training leads to significant gains but further increase the sample ratio results in limited improvements. }}
\label{fig4}
% \vspace{-0.4cm}
\end{figure}
\subsection{Offline Performance}
Table \ref{overall} presents the comprehensive comparison between OneRec and various baselines. For watching-time metric we mainly care about the session watch time (\textbf{swt}) and like probability (\textbf{ltr}) in interaction metrics. Our result reveals three key observations:

\textbf{First, the proposed session-wise generation approach significantly outperforms traditional dot-product-based methods and point-wise generation methods like TIGER.} OneRec-1B achieves 1.78\% higher maximum swt and 3.36\% higher maximum ltr compared to TIGER-1B. This demonstrates the advantage of session-wise modeling in maintaining contextual coherence across recommendations, whereas point-wise methods struggle to balance coherence and diversity in generated outputs.

\textbf{Second, a small ratio of DPO training yields substantial gains.} With only 1\% DPO training ratio ($r_{\rm DPO}$), OneRec-1B+IPA surpasses the base OneRec-1B by 4.04\% in maximum swt and 5.43\% in maximum ltr. This suggests limited DPO training can effectively aligns the model with desired generation patterns. 

\textbf{Third, the proposed IPA strategy outperforms various existing DPO variants.} As shown in Table \ref{overall}, IPA achieves superior performance compared to alternative DPO implementations. Notably, some DPO baselines underperform even the non-aligned OneRec-1B model, suggesting that iterative mining of self-generated outputs for preference selection proves more effective than other methods.
\begin{figure*}[hbpt]
\centering
\includegraphics[width=.99\textwidth]{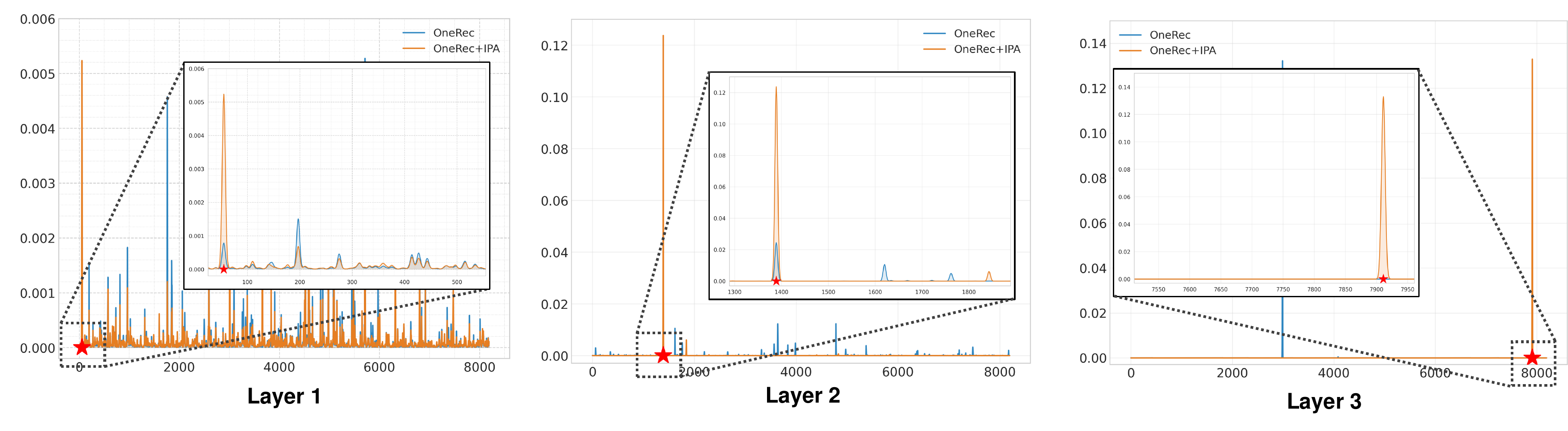}
\caption{\textbf{The visualization of the probability distribution of the softmax output for each layer of the semantic ID. The red star represents the sematic ID of item which has the highest reward value.}}
\label{fig6}
% \vspace{-0.4cm}
\end{figure*}

\begin{figure}[hbpt]
\centering
\includegraphics[width=.49\textwidth]{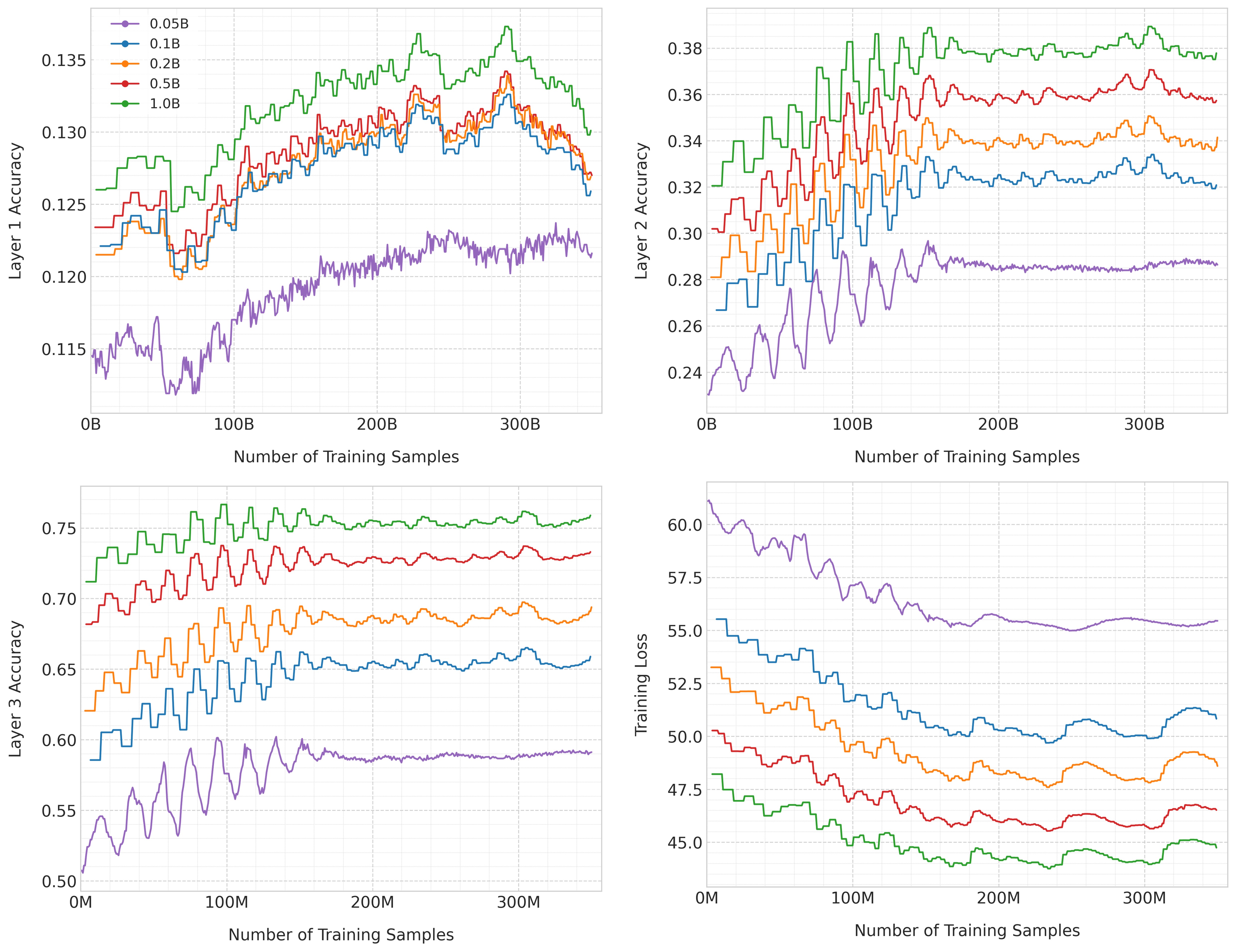}
\caption{\textbf{Scalability of OneRec on model scaling. The results show that OneRec constantly benefits from performance improvement when the parameters are scaled up.}}
\label{fig5}
\vspace{-0.2cm}
\end{figure}
\vspace{-0.3cm}
\subsection{Ablation Study} 
\subsubsection{DPO Sample Ratio Ablation} In order to investigate the impact of sample ratio $r_{\rm DPO}$ in DPO training, we varied the DPO sample ratio from 1\% to 5\% under controlled conditions. As illustrated in Figure \ref{fig4}, ablation results demonstrate that increasing the sample ratio yields marginal performance improvements across multiple evaluation targets. Notably, the performance gains beyond the 1\% baseline remain insignificant despite increased computational expenditure. It worth noting that there exists a linear relationship between and GPU resource utilization during DPO sample server inference: the 5\% sample ratio requires $5 \times$ more GPU resources than the 1\% baseline. This scaling characteristic establishes an explicit trade-off between computational efficiency and model performance. Therefore, after balancing the best trade-off with computation efficiency and performance, we apply 1\% DPO sample ratio for training, which achieves average 95\% of the maximum observed performance while requiring only 20\% of the computational resources needed for higher sample ratio.

\subsubsection{Model Scaling Ablation} We evaluate how OneRec performs when the model scale increases. As Figure \ref{fig5} shows, scaling OneRec from 0.05B to 1B achieves consistent accuracy gains, demonstrating consistent scaling properties. Specifically, compared to OneRec-0.05B, OneRec-0.1B achieves a significant maximum 14.45\% gain in accuracy, and 5.09\%, 5.70\% and 5.69\% additional accuracy gains can be achieved when scaling to 0.2B, 0,5B and 1B.

\vspace{-0.2cm}
\subsection{Prediction Dynamics of OneRec} 
As shown in Figure \ref{fig6}, we present the predicted probability distributions of 8192 codes across different layer, where the red star denotes the semantic ID of the item with the highest reward value. Compared to the OneRec baseline, OneRec+IPA exhibits a significant confidence shift in prediction distributions, indicating that our proposed preference alignment strategy effectively encourages the base model to produce preferred generation patterns. Furthermore, we observe that the probability distribution in the first layer demonstrates greater divergence (entropy = 6.00) compared to subsequent layers (average entropy = 3.71 in the second layer and entropy = 0.048 in third layer), which exhibit progressively concentrated distributions. This hierarchical uncertainty reduction can be attributed to the autoregressive decoding mechanism: the initial layer's predictions inherit higher uncertainty from preceding decoding steps, while later layers benefit from accumulated context that constrains the decision space.

\subsection{Online A/B Test}
To evaluate the online performance of OneRec, we conduct strict online A/B tests on Kuaishou’s video recommendation scenarios of main page and we compare the performance of OneRec and current multi-stage recommender system with 1\% main traffic for experiments. We use \textit{Total Watch Time} to measure the total time that users spend watching videos and \textit{Average View Duration} calculates the average watch time per video when the user is exposed to a requested session by the recommendation system. Online evaluation shows that OneRec has achieved \textbf{1.68\%} improvement in total watch time and \textbf{6.56\%} improvement in average view duration, which indicates that OneRec achieves much better recommendation results and brings considerable revenue increments for the platform.
\begin{table}[htb]
    \centering
    \caption{The absolute improvement of OneRec compared to the current multi-stage system in the online A/B testing setting.}
    \resizebox{0.80\linewidth}{!}{
    \begin{tabular}{
    >{\centering\arraybackslash}p{2.5cm}  % Method列
    >{\centering\arraybackslash}p{2.5cm}    % Avg-Play
    >{\centering\arraybackslash}p{3.0cm}    % Click
}
    \toprule
        Model & Total Watch Time & Average View Duration \\
    \midrule
         OneRec-0.1B & +0.57\%  & +4.26\% \\
         OneRec-1B& +1.21\%   & +5.01\%\\
        OneRec-1B+IPA & +1.68\%  & +6.56\%\\ 
    \bottomrule
    \end{tabular}
    }
    \label{ab_test}
    \vspace{-0.3cm}
\end{table}

\section{Conclusion}
In this paper, we focus on the introduction of an industrial solution for single-stage generative recommendation. Our solution establishes three key contributions: First, we effectively scale the model parameters with high computational efficiency by applying the MoE architecture, offering a scalable blueprint for large-scale industrial recommendation. Next, we find the necessity of modeling the contextual information of target items in a session-wise generation manner, proving contextual sequence modeling inherently captures user preference dynamics better than isolated point-wise manner. Furthermore, we propose an Iterative Preference Alignment (IPA) strategy to improve OneRec's generalization across diverse user preference patterns. Extensive offline experiments and online A/B testing verify the effectiveness and efficiency of OneRec. Additionally, our analysis of online results reveals that, besides user watch time, our model has limitations in interactive indicators, such as likes. In future research, we aim to enhance the end-to-end generative recommendation's capability in multi-objective modeling to provide a better user experience.

\newpage
\bibliographystyle{ACM-Reference-Format}

\bibliography{onerec}
\end{document}